\documentclass[twocolumn,showpacs,preprintnumbers,amsmath,amssymb,floatfix]{revtex4}
\usepackage{graphicx}
\usepackage{dcolumn}
\usepackage{bm}

\newcommand{\comment}[1]{}

\newcommand{\vrms}{\Delta v}

\begin{document}

\title{Coherence-enhanced imaging of a degenerate Bose gas}

\author{
\ L. E. Sadler, J.\ M.\ Higbie, S. R.\ Leslie, M. Vengalattore, and
D.\ M.\ Stamper-Kurn} \affiliation{Department of Physics, University
of California, Berkeley CA 94720}

\date{\today }

\begin{abstract}
We present coherence-enhanced imaging, an \textit{in situ} technique
that uses Raman superradiance to probe the spatial coherence
properties of an ultracold gas.  Applying this method, we obtain a
spatially resolved measurement of the condensate number and more
generally, of the first-order spatial correlation function in a gas
of $^{87}$Rb atoms.  We observe the enhanced decay of propagating
spin gratings in high density regions of a Bose condensate, a decay
we ascribe to collective, non-linear atom-atom scattering.  Further,
we directly observe spatial inhomogeneities that arise generally in
the course of extended sample superradiance.
\end{abstract}


\maketitle

Ultracold Bose and Fermi gases represent new condensed-matter
systems in which to study the conditions for and onset of long-range
off-diagonal order\ \cite{grei02mott}. Beyond serving as models of
existing quantum fluids, gaseous systems allow for studies of
coherent matter far from equilibrium.  Studying such non-equilibrium
samples may yield insights into symmetry-breaking dynamics at
thermal and quantum phase transitions\
\cite{zure05qpt,kibb76,zure85cosmo}.  A direct probe of off-diagonal
order in these dilute gases is essential to fully exploit this
opportunity.

Limitations of existing techniques for measuring coherence have
narrowed the scope of prior studies. In time-of-flight methods,
coherence is detected through matter-wave interference, which
provides poor spatial resolution. This reduced resolution may
preclude studying coherence in inhomogeneous systems such as at the
borders of Mott-insulating regions of bosons in optical lattices
\cite{jaks98lattice}, gases trapped in disordered potentials\
\cite{lye05}, or systems undergoing a first-order phase transition
to a superfluid state
\cite{altm03twocomp,krut04qpt,kukl04first,kimu05first}.  A second
technique, the imaging of irrotational flow as a signal for phase
coherence, requires the gaseous system to be strongly perturbed over
extended periods of time. This agitation prevents direct
measurements of phase coherence in rapidly evolving systems or in
systems near a phase transition.

In this Letter, we describe coherence-enhanced imaging, which allows
the coherent portions of a gaseous system to be directly identified
with high spatial resolution. Previous \textit{in situ} imaging
methods \cite{andr96,andr97prop,hau98}, being sensitive to the
linear optical susceptibility of the trapped gas, allow the
identification of the coherent portion of the gas only through its
high optical density.  In contrast, our imaging method relies on a
non-linear optical property of the trapped gas,
superradiance-enhanced absorption, to gain access to the difference
in \emph{coherence} between the condensed and non-condensed portions
of the gas.

We present two main results derived from coherence-enhanced images.
First, we quantify coherence properties in a degenerate $^{87}$Rb
gas.  Specifically, we obtain a spatially resolved measure of the
condensate number without resorting to model-dependent fitting of
spatial images\ \cite{kett99var}.  We measure the first-order
spatial correlation function in a partly condensed gas with high
spatial resolution. Second, we obtain the first spatially and
temporally resolved study of extended sample superradiance,
revealing inhomogeneous collective scattering across the gas.

Collective light scattering has been observed in many systems
including solid crystals\ \cite{flor84}, plasmas\ \cite{dreh04}, and
gases\ \cite{skri73}.  Such scattering can be regarded as a form of
extended-sample superradiance \cite{dick54} where the pump light
that illuminates the sample places all scatterers into an excited
state from which they may optically decay.  The light emitted from
these systems is highly directed and is the primary signal used to
study the temporal evolution of this process. An additional
signature of superradiance in ultracold gases is the directed
emission of atoms from the original atomic sample following the
momentum recoil from superradiant light emission\
\cite{inou99super2}. Here we focus on a new signature of
superradiance, namely the enhanced absorption of the pump light due
to collectively enhanced light scattering. Imaging this light yields
detailed spatial information on the radiating sample.

Spontaneous collective light scattering is most easily described in
the context of Rayleigh superradiance.  In this process,
off-resonant light pumps atoms to an excited state whence they
decay, recoiling with momentum $\hbar \mathbf{q}=\hbar
\mathbf{k}_{i}-\hbar \mathbf{k}_s$ which is the difference between
the momentum of the incident and scattered photons.  Spatial
coherence between the recoiling and stationary atoms causes a
periodic density grating to form through matter-wave interference.
 This density grating acts as a partially reflective mirror, oriented so as to scatter light
 in the same direction as previously scattered light, and the reflectivity of which
 is proportional to the local density of unscattered and scattered
 atoms.
 In a prolate sample, the enhanced light scattering is directed
 predominately along the end fire modes (EFM), i.e.\ along the long
 axis of the gas.  The mechanism for Raman superradiance, which is used in this work,
differs in that interference between recoiling and stationary atoms
creates a spin-polarization rather than a number-density grating and
in that recoiling atoms no longer scatter light\
\cite{schn04raman,yosh04}.

 The enhancement of the optical scattering rate depends solely on the
quality of the grating formed in the gas.  In turn, properties of
the grating depend on the competition between a position-dependent
loss of grating coherence and the amplification of the grating by
stimulated emission.  The loss of grating coherence can occur due to
Doppler dephasing or collisional dephasing and loss. Considering
just Doppler dephasing, the polarization grating formed by the
unscattered and recoiling atoms will decay on a timescale $\tau_c
\sim (q \vrms)^{-1}$ with $\vrms$ being the local atomic rms
velocity. This correlation time, $\tau_c$, is the time it takes
atoms propagating at momentum $\hbar \mathbf{q}$ to travel beyond
the coherence length $\lambda = \hbar / m \vrms$, with $m$ being the
atomic mass. As a result, polarization gratings in gases with long
coherence lengths, such as a Bose-Einstein condensate (BEC), persist
much longer than those in non-degenerate gases with short,
temperature ($T$) dependent coherence lengths $\lambda_{dB} =
\sqrt{2 \pi \hbar^2 / m k_B T}$.

In our experiment, we induce superradiance in a gas using multiple,
short pulses of light separated by a variable delay time $\tau$.
Enhanced absorption, which we image, is built up over multiple
pulses in the portion of the gas with $ \tau\ll\tau_c$, while
Doppler dephasing in the portion of the sample with $\tau\gg\tau_c$
suppresses the pulse-to-pulse growth of the polarization grating.
Our imaging method provides a spatially resolved measure of the
first-order spatial correlation function. That is, an optical
absorption signal is obtained that quantifies the modulation depth
of interference between the stationary gas and its replica,
displaced by $\delta r=\hbar q\tau/m$.

\begin{figure}
\includegraphics[width=0.45\textwidth]{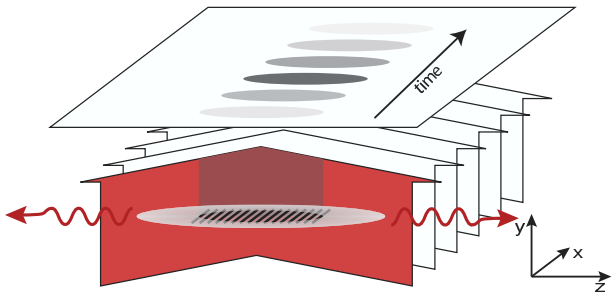}
 \caption{Experimental scheme for coherence-enhanced imaging.
  Short-pulses of light illuminate a prolate gas perpendicular
  to the magnetic field axis ($\hat{z}$).  Superradiance establishes
   a coherent polarization grating in spatially coherent portions
  of the gas, causing collective absorption of probe light and
 its re-emission along the end-fire modes.  Each incident pulse
 is imaged separately, yielding images such as shown in Figure 2.
 }
 \label{fig:scheme}
\end{figure}

Our experiments were performed on anisotropic (cigar-shaped) gases
of $^{87}$Rb trapped in a Ioffe-Pritchard magnetic trap in the
$|F=1, m_F = -1\rangle$ hyperfine state. The magnetic trap was
characterized by trap frequencies of $\omega_{x,y,z} = 2 \pi
(48,48,5)\,\mbox{s}^{-1}$ in the transverse ($\hat{x}$,$\hat{y}$)
and axial ($\hat{z}$) directions.  The gas temperature was varied by
the final settings used in rf evaporative cooling, yielding gases of
$15 \times 10^6$ atoms at the BEC transition temperature of 250 nK,
and pure condensates of up to $1.6 \times 10^6$ atoms.

The light used for superradiance-enhanced imaging was directed along
the $\hat{y}$ axis, linearly polarized perpendicular to the
$\hat{z}$ magnetic field axis (Fig.\ \ref{fig:scheme}), and detuned
by 102 MHz above the $F$=1$ \rightarrow$$ F^\prime$=1 D1 transition
($k_i = 2 \pi / 795 \, \mbox{nm}$). This detuning and polarization
was chosen for two reasons. First, Rayleigh scattering is eliminated
due to destructive interference among the transitions to the
$F^\prime$$ = 1$ and the $F^\prime$$=2$ excited hyperfine states.
This eliminates the dispersive phase shift that would otherwise
cause aberrations in imaging the  $\sim\!10\, \mu$m narrow cloud.
 Second, at this detuning, effects on collective light scattering
 from
 both Zeeman and mean-field interaction shifts are made negligible due to
 a predominant superradiant scattering to the $|F=2, m_F = 1\rangle$
 hyperfine state.  This final state possesses a nearly equal magnetic
 moment to the initial state, and the
 relative $s$-wave scattering lengths among all trapped states are
 nearly identical.

\begin{figure}
\centering
 \includegraphics[width=0.45\textwidth]{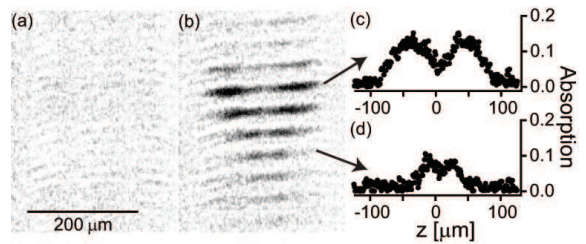}
  \caption{Direct, \textit{in situ} imaging of coherence in a BEC.
  Coherence-enhanced absorption images are shown for single trapped Bose
  gases
 at (a) $T/T_c$=1.3 and (b) $T/T_c$=0.3.  Each of ten frames is illuminated by a 100 $\mu$s pulse of light separated by
  $\tau=68\ \mu$s delay times and is shown with time evolving from the
 top to the bottom frame.  Doppler dephasing during the delay times suppresses
 superradiance except in coherent portions of the gas.  Thus, enhanced absorption
 is seen only in the condensed portions of the degenerate sample (b).  The absorption signal varies
  across the cloud with collective enhancement occurring first at the ends of the extended sample.}
    \label{fig:timeresolve}
\end{figure}

Coherence-enhanced images of a pure condensate taken at this
detuning (Fig.\ \ref{fig:timeresolve}) demonstrate the spatial and
temporal resolution of our technique.  For these images, a BEC of
1.6$\times$10$^6$ atoms was illuminated by a series of 100
$\mu$s-long pulses of light, with a $\tau = 68 \, \mu$s delay
between pulses. The transmitted part of each probe pulse was
separately imaged at a diffraction-limited resolution of  $\sim 6\,
\mu$m. The absorption by the condensate increases over several probe
pulses due to the collective Raman scattering culminating in a peak
absorption of 15\%.  After achieving maximum enhancement, the
absorption diminishes as a large fraction of the condensed atoms
have been optically pumped to the $|F=2, m_F=1\rangle$ state in
which they no longer absorb probe light.

To estimate the maximum absorption expected for this sample, let us
consider superradiant scattering solely into one EFM and into one
final atomic state ($|F=2,m=1\rangle$).  In the case when half the
atoms have been scattered and presuming simultaneous collective
behavior by all atoms in the sample, the optical depth is enhanced
by a factor $\mathcal{E} = \frac{R_j}{R_{tot}}\frac{ f_j}{4} N_0
A_\Omega$.
 Here, $R_j/R_{tot}=0.28$ is the branching ratio for scattering into the preferred final state
$|j\rangle$, $f_j$ is an angular term that includes the dipole
emission pattern, $N_0$ is the total atom number, and $A_\Omega$ is
a phase-matching integral\ \cite{inou99super2}; here $N_0 A_\Omega =
5800$.  We thus predict a maximum enhancement factor $\mathcal{E}
\simeq 60$, somewhat greater than the observed $\mathcal{E} \simeq
10$.

This discrepancy in enhancement is not surprising given the
significant spatial features seen in the coherence-enhanced images
(Fig.\ \ref{fig:timeresolve}).  These images show a distinctive
pattern in which the ends of the extended sample darken earlier than
the denser cloud center.  Thus, the maximal enhancement of light
scattering does not occur simultaneously throughout the cloud,
diminishing the overall enhancement of absorption as compared with
the estimate described above.  This diminished enhancement is most
pronounced at the center of the gas because the majority of atoms in
the sample have been transferred to the recoiling state, diminishing
the depth of the grating from which light is scattered.

 Extensions to the simplest
superradiance models have been constructed \cite{macg76,Uys} that
predict the spatial inhomogeneities that we observe.  In these
models, the initiation of collective scattering is treated in a
manner akin to Dicke's original work\ \cite{dick54} in which light
scattering leads to the occupation of atomic recoil modes that are
propagating replicas of the unscattered atomic state.  Once
collective scattering dominates, subsequent developments are
described semi-classically. Here, light propagating down the long
axis of the sample acquires its highest intensity at the tip of the
gas.  Atoms located at the tips are thus more strongly stimulated by
the field and emit light earlier than those in the center of the
cloud. While the development of spatial structure was indicated
indirectly by previous studies\ \cite{schn03amp,yosh05super},
coherence-enhanced imaging gives unprecedented temporal and spatial
resolution of the entire superradiant process.

\begin{figure}
\centering
\includegraphics[width=0.45\textwidth]{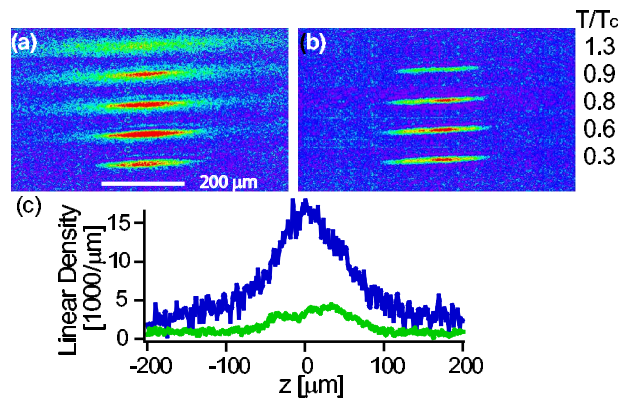}
  \caption{Spatial maps of condensate number.  (a) Dispersion-free off-resonant linear Raman
  absorption images are given for varying $T/T_c$ revealing the
  combined densities of the condensed and normal fractions of the
  gas, while providing no direct information on its coherence.  (b)  Coherence-enhanced images, derived from
  collating many frames from a multiple pulse imaging sequence, exclusively reveal
  portions of the gas that are phase-coherent.  (c)
  Radially summed (across $\hat{x}$) cross-section for a gas at $T/T_c$=0.9 shows a  bimodal
  (blue) total number
  distribution in linear absorption imaging and condensate-only distribution (green)
  in coherence-enhanced images, which is 25$\%$ of the total number. }
      \label{fig:conmap}
\end{figure}

Beyond providing insight on the dynamics of superradiance, these
multiple-frame images yield quantitative information about spatial
coherence in trapped gases.  For example, in contrast with linear
absorption imaging, coherence-enhanced imaging allows the coherent
fraction of the partially condensed Bose gas to be spatially mapped
and its population counted.  Images obtained by each of these
methods are compared in Fig.\ \ref{fig:conmap}. The off-resonant
linear absorption images, taken with light 200 MHz below the D2
$F$=1$\rightarrow$$F$$'$=2 transition to eliminate dispersive phase
shifts, show a typical progression of density distributions from
Gaussian for $T/T_c>1$, to bimodal, and finally to a single
parabolic (condensate) density distribution for the smallest
$T/T_c$. In contrast, coherence-enhanced absorption appears only
when $T/T_c<$1.  Here, we present maps, derived from multiple-frame
images such as those in Fig.\ \ref{fig:timeresolve}, that give the
net coherence-enhanced absorption, i.e.\ each map pixel shows the
total number of missing probe photons summed over all image frames.
Coherence enhanced imaging selectively resolves the condensed
portion of the gas, whereas this portion is obscured in the linear
absorption images. Under the assumption that the condensate is
completely optically pumped to the $|F=2, m_F = 1\rangle$ state
during multiple-pulse superradiance, the number of missing photons
directly quantifies the condensate number.

Moreover, by varying the delay times $\tau$ in our imaging sequence,
we obtain a spatially resolved measurement of the first order
spatial correlation function.  For this purpose, we determine, at
each pixel location, the ratio of absorption between the first and
second frame in the image. This technique, similar to that of Ref.\
\cite{yosh05super} in which the ratio of superradiant light emission
intensities was considered, measures the spatial correlations
between atoms at varying distances, $\delta r=\hbar q\tau/m$. The
absorption ratio, which measures the overlap of the recoiling and
stationary spatial wavefunctions, is large in portions of the gas
with high spatial coherence and tends to unity when the correlations
being probed exceed the local coherence length.

\begin{figure}
\centering
 \includegraphics[width=0.45\textwidth]{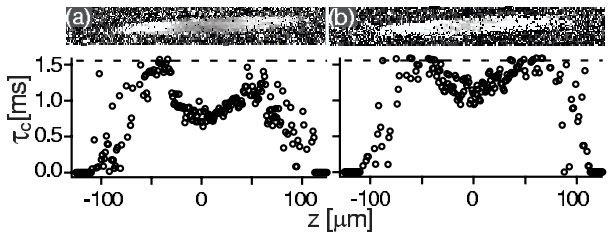}
  \caption{Maps of correlation times $\tau_c(x,z)$ shown for degenerate Bose gases at
  (a) $T/T_c=0.3$ and (b) $T/T_c$=0.8.  Longitudinal cross-sections are shown below the corresponding images.
    The Doppler limit on the correlation time (dashed line)
  matches measurement at the least dense but fully coherent tips of
  the condensate.  The reduction of $\tau_c$ at the condensate center is ascribed to non-linear
  atom-atom elastic scattering.  Gray scale runs from 0 to 1.6 ms.}
    \label{fig:length}
\end{figure}

Multiple-pulse coherence-enhanced images were taken of identically
prepared gaseous samples with variable delay times $\tau$.  In Fig.\
\ref{fig:length} we present maps of the characteristic coherence
time $\tau_c(x,z)$, which we obtain at each pixel as the $1/e$ decay
time of the absorption ratio as determined by gaussian fits.  We
observe a maximal coherence time, $\tau_c=1.6$ ms, which is
consistent with the time required for recoiling atoms to travel
through the condensate. This shows that BECs are coherent across
their radial width as previously seen in Ref.\
\cite{andr97int,sten99brag,hagl99coh}.

Surprisingly, however, we find that $\tau_c$ diminishes in the
center of the BEC, most dramatically for samples at the lowest
temperatures at which the condensate fraction and density are the
highest.  Rather than taking this observation to imply a reduced
spatial coherence at the center of the condensate, we ascribe the
lowering of $\tau_c$ to non-linear collisional depletion of the
propagating spin grating formed in superradiance\ \cite{chik00}. The
linear (non-collective) per-particle scattering rate of atoms
recoiling at $v_{rec}$ propagating through the stationary portion of
the gas, $\Gamma_l=n\sigma v_{rec}= 300$ s$^{-1}$ is too small to
significantly affect the observed correlation times.  However, in a
coherent gas, such scattering is enhanced to the extent that the
total number of collisions, $N_s$ exceeds the number of quantum
states $M$ available to the collision products. We estimate the
enhanced non-linear scattering rate in a volume with a radius that
is equal to the distance the sample recoils before the grating
decays to be $\Gamma_{nl}=\sqrt{(\Gamma_l v_{rec}\lambda^2
n_s)/6\pi} \simeq$ 1/(500 $\mu$s) where $\lambda=795$\ nm/$\sqrt{2}$
and $n_s$ is the density of the scattered atoms.  This rate is in
good agreement with the value of $\tau_c$ observed at the center of
the condensate. In more dilute regions, such as the tips of the
condensate, or in the center of less dense condensate formed at
higher temperatures, the collective enhancement of collisional
losses is weaker, and thus, higher values of $\tau_c$ are observed.

As we have shown, coherence-enhanced imaging yields simultaneous
information both on the spatial coherence of an ultracold gas and
also on superradiance in extended samples.  Quantitative measures of
the first-order spatial correlation function were obtained at high
spatial resolution over the extent of the gas. However, the spatial
structures seen to appear in the course of superradiance are an
inconvenient feature that complicate the analysis. Alternatively,
one could employ a two-photon stimulated Raman transition to set up
the initial, now uniform, polarization grating, and then image the
collectively enhanced absorption of a subsequent single probe beam.
Such an approach is technically complicated by the need for two
laser beams with a precise frequency difference and also by
sensitivity to overall Doppler shifts for a possibly moving gas
sample.  Our method avoids such complications.

On the other hand, coherence-enhanced imaging as implemented here
has allowed for the first direct temporal and spatial study of the
onset of inhomogeneous collective scattering and offers avenues to
further study.  As pointed out recently\ \cite{Uys}, fluctuating
asymmetries in the spatial structure that develop spontaneously
during superradiance may be studied to characterize further the
quantum fluctuations that play a role in its early stages.

We thank P. Meystre and H. Uys for discussions and D. Schneble for
comments on the manuscript.  This work was supported by the NSF and
the David and Lucile Packard Foundation. S.R.L.\ acknowledges
support from the NSERC.

\bibliographystyle{apsrev}
\comment{\bibliography{allrefs,sr}}

\end{document}